\documentclass[letterpaper, 10 pt, conference]{ieeeconf}  

\IEEEoverridecommandlockouts                              
\usepackage{amsmath,amsthm,amssymb,mathtools,bbm,dsfont,physics,aligned-overset}
\allowdisplaybreaks

\usepackage[hidelinks]{hyperref}
\usepackage[capitalize]{cleveref}

\usepackage{xcolor}
\definecolor{myorange}{RGB}{255,170,34}
\definecolor{ETHlightblue}{RGB}{0, 122, 150}
\definecolor{ETHpurple}{RGB}{145, 5, 106}
\definecolor{ETHgray}{RGB}{111, 111, 111}
\definecolor{ETHblue}{RGB}{0, 105, 180}
\definecolor{ETHred}{RGB}{168, 50, 45}

\usepackage{comment}

\usepackage{tikz,pgfplots,pgfplotstable}
\usetikzlibrary{calc,external}
\usepgfplotslibrary{groupplots,statistics}
\tikzset{font=\footnotesize}
\pgfplotsset{compat=1.16}
\pgfplotsset{
    colormap={mycolorbar}{color=(ETHblue); color=(ETHred)} 
}

\usepackage{microtype}

\usepackage{etoolbox}
\newbool{booltrue} 
\booltrue{booltrue}
\newbool{boolfalse} 
\boolfalse{boolfalse}

\usepackage{cite}

\usepackage{caption,subcaption}
\AtEndEnvironment{figure}{\vspace{-0.15cm}}
\AtEndEnvironment{figure*}{\vspace{-0.15cm}}

\newtheorem{theorem}{Theorem}
\newtheorem{lemma}[theorem]{Lemma}

\newtheorem{proposition}[theorem]{Proposition}

\theoremstyle{definition}

\newtheorem{example}{Example}
\newtheorem*{remark}{Remark}

\newcommand{\reals}{\mathbb{R}}
\newcommand{\naturals}{\mathbb{N}}
\newcommand{\nonnegativereals}{\reals_{\geq 0}}
\renewcommand{\d}{\mathrm{d}}
\DeclareMathOperator*{\argmax}{argmax}

\newcommand{\wassersteinLetter}[1]{W_{#1}}
\newcommand{\wasserstein}[3]{\wassersteinLetter{#1}(#2,#3)}
\newcommand{\wassersteinLarge}[3]{\wassersteinLetter{#1}\left(#2,#3\right)}
\newcommand{\pushforward}[1]{#1_{\#}}
\newcommand{\Pp}[2]{\mathcal{P}_{#1}(#2)}
\newcommand{\setPlans}[2]{\Gamma(#1,#2)}

\DeclareMathOperator*{\variance}{Var}
\newcommand{\expectedValue}[2]{\mathbb{E}^{#1}[#2]}

\newcommand{\opinion}       {x}
\newcommand{\dopinion}      {\mu}
\newcommand{\rec}           {u}
\newcommand{\drec}          {\rho}
\newcommand{\weightopinion} {\beta}
\newcommand{\weightbias}    {\alpha}
\newcommand{\exploration}   {T}
\newcommand{\reward}        {r}

\newbool{finalsubmission} 
\boolfalse{finalsubmission} 

\title{\LARGE \bf
The Impact of Recommendation Systems on Opinion Dynamics: Microscopic versus Macroscopic Effects
}

\author{Nicolas Lanzetti, Florian Dörfler, and Nicolò Pagan
\thanks{This work was supported by the Swiss National Science Foundation under NCCR Automation, grant agreement 51NF40\_180545.}
\thanks{N. Lanzetti and F. Dörfler are with the Automatic Control Laboratory, Department of Electrical Engineering and Information Technology, ETH Zürich, 8003 Zürich,  Switzerland, {\tt\small \{lnicolas,doerfler\}@ethz.ch}}%
\thanks{N. Pagan is with the Social Computing Group, Department of Informatics, University of Zürich, 8003 Zürich, Switzerland, {\tt\small nicolo.pagan@uzh.ch}}%
}

\begin{document}

\maketitle
\thispagestyle{empty}
\pagestyle{empty}

\begin{abstract}
Recommendation systems are widely used in web services, such as social networks and e-commerce platforms, to serve personalized content to the users and, thus, enhance their experience.
While personalization assists users in navigating through the available options, there have been growing concerns regarding its repercussions on the users and their opinions. Examples of negative impacts include the emergence of filter bubbles and the amplification of users' confirmation bias, which can cause opinion polarization and radicalization.
In this paper, we study the impact of recommendation systems on users, both from a microscopic (i.e., at the level of individual users) and a macroscopic (i.e., at the level of a homogenous population) perspective. Specifically, we build on recent work on the interactions between opinion dynamics and recommendation systems to propose a model for this closed loop, which we then study both analytically and numerically. 
Among others, our analysis reveals that shifts in the opinions of individual users do not always align with shifts in the opinion distribution of the population. In particular, even in settings where the opinion distribution appears unaltered (e.g., measured via surveys across the population), the opinion of individual users might be significantly distorted by the recommendation system.

\end{abstract}


\section{Introduction}\label{sec:introduction}

Over the past few years, recommendation systems have become an essential component of online services, including e-commerce platforms and social networking sites. Their primary objective is to filter through the vast amount of information available and guide users towards the most relevant content.
Recommendation systems make use of diverse machine learning methods to assess the relevance of items and provide personalized content based on the recorded online behaviors of users. 
These techniques enable the systems to not only measure the absolute relevance of items but also tailor the recommendations to the users' expected tastes~\cite{konstan2012recommender}.
While recommendation systems have been a remarkable technological advancement, their impact on users' behavior has raised questions. 
Personalization is a key feature of these systems that improves the user experience but also poses concerns: Excessive personalization may limit the range of perspectives available to users, leading to ``filter bubbles''~\cite{pariser2011filter}. These bubbles can induce opinion polarization and radicalization, which can be harmful~\cite{sunstein2018republic}. 
Although later research~\cite{bakshy2015exposure} has downplayed concerns about the negative effects of personalization, evidence suggests that it has the potential to strengthen users' prejudices. For instance, numerous studies showed that it aggravates confirmation bias, which is the human propensity to seek and consider information that confirms their beliefs and ideas~\cite{nickerson1998confirmation,klayman1987confirmation}. This bias can lead to an unconscious one-sided argument-building process, reinforcing users' preconceived notions. Therefore, it is reasonable to conclude that personalization may exacerbate the confirmation bias phenomenon, potentially leading to further polarization and division among users.

Since empirical evidence supports the idea that confirmation bias is extensive, strong, and multiform, and its effects may be amplified by curation algorithms~\cite{sunstein2018republic}, a recent stream of literature~\cite{pagan2023classification,sinha2016deconvolving, perra2019modelling, rossi2021closed, dean2022preference} has started exploring the impact of the closed-loop dynamics between personalized recommendations and user preferences and opinions.
For example, \cite{perra2019modelling,rossi2021closed} examined how this loop can reinforce user preferences and lead to polarization and filter bubbles, with \cite{rossi2021closed} focusing mainly on the interaction between the user and the recommendation system, while \cite{perra2019modelling} also included the effect of social influence by considering users being embedded in their social network.
Differently,~\cite{sinha2016deconvolving, dean2022preference} studied how to disentangle feedback loops in order to improve recommendation accuracy.
The overall goal of this research field is to overcome the potential negative consequences of personalization by designing recommendation algorithms that can influence users' opinions and preferences in a more beneficial way~\cite{lerman2010information,anderson2013steering}.

While these works generally show that the closed loop between opinion dynamics and recommendation systems bears the potential to steer individuals' opinions, it remains unclear (i) to what extent it may also steer the opinion distribution of a population (thus leading to significant concerns with regard to, e.g., political debates) and (ii) if seemingly similar opinion distributions (at the population level, e.g., surveys) can hide substantial individual shifts.
To shed light on these questions, this paper builds on the recent model of~\cite{rossi2021closed} and examines the micro- and macroscopic impact of the closed loop between users and a recommendation system. This way, we can both study the impact of recommendation systems on the opinion distribution of a population and determine if shifts in opinions at the individual level can be concealed by their cumulative effect on the population level.

\paragraph*{Contributions}
With our work, we formalize the interaction between users and recommendation systems both from a microscopic (i.e., one user interacting with a recommendation system) and macroscopic (i.e., a homogeneous population interacting with a recommendation system) perspective. 
This way, we identify several tractable yet insightful instances, which we can investigate analytically and help us shed light on the impact of recommendation systems on users' opinions. Among others, our analytical analysis and numerical simulations uncover and explain a discrepancy between micro- and macroscopic behaviors, whereby the opinion of individual users is highly impacted by the recommendation system while, macroscopically, the opinion distribution remains unaffected. 
This insight reveals that, even when population surveys (e.g., exit polls) do not indicate opinion shifts, individuals' beliefs might be highly impacted by the recommendation systems.

\paragraph*{Organization}
This paper unfolds as follows.
In~\cref{sec:model}, we present our model of the closed loop between a user and a recommendation system. We study its properties in~\cref{sec:analysis} and present numerical results in~\cref{sec:results}. \cref{sec:conclusion} draws the conclusions of this paper. \ifbool{finalsubmission}{Proofs are included in the online extended version of this paper.}{Proofs are deferred to the appendix.} 

\subsection{Notation and background material}
We denote by $\nonnegativereals$ the non-negative real numbers. The space of (Borel) probability distributions over $\reals$ is $\Pp{}{\reals}$ and the space of probability distributions over $\reals$ with finite second moment is $\Pp{2}{\reals}\coloneqq \{\mu\in\Pp{}{\reals}:\int_{\reals}\abs{x}^2\d\mu(x)<+\infty\}$.
The mean of a probability distribution $\mu\in\Pp{}{\reals}$ is $\expectedValue{\mu}{x}$ and its variance is $\variance(\mu)$.
The pushforward of a probability distribution $\mu\in\Pp{}{\reals}$ via a (Borel) map $f:\reals\to\reals$ is defined by $(\pushforward{f}\mu)(A)=\mu(f^{-1}(A))$ for each Borel set $A\subset\reals$; if $X\sim\mu$, then $Y=f(X)\sim\pushforward{f}\mu$.
The convolution of two probability distributions $\mu,\nu\in\Pp{}{\reals}$ is denoted by $\mu\ast\nu$; if $X\sim\mu$ and $Y\sim \nu$ are independent, then $X+Y\sim\mu\ast\nu$.
The (type-$p$) Wasserstein distance between two probability distributions $\mu,\nu\in\Pp{}{\reals}$ is defined by
\begin{equation*}
    \wasserstein{p}{\mu}{\nu}
    \coloneqq
    \left(\min_{\gamma\in\setPlans{\mu}{\nu}}\int_{\reals\times\reals}\abs{x-y}^p\d\gamma(x,y)\right)^{\frac{1}{p}},
\end{equation*}
where $\setPlans{\mu}{\nu}\subset\Pp{}{\reals\times\reals}$ is the set all probability distributions over $\reals\times\reals$ with marginals $\mu$ and $\nu$ (referred to as transport plans)~\cite{villani2009optimal}.
The Wasserstein distance is the minimum cost to transport $\mu$ onto $\nu$ when transporting a unit of mass from $x$ to $y$ costs $\abs{x-y}^p$. Accordingly, a transport plan $\gamma\in\setPlans{\mu}{\nu}$ encodes the allocation of probability mass: If $(x,y)$ is in the support of $\gamma$, then some of the probability mass at $x$ is displaced to $y$, or, equivalently, $\gamma(A\times B)$ is the mass transferred from the set $A\subset\reals$ to the set $B\subset\reals$.
Finally, a sequence of probability distributions $(\mu_n)_{n\in\naturals}\subset\Pp{2}{\reals}$ converges weakly in $\Pp{2}{\reals}$ if $\int_{\reals}\phi(x)\d\mu_n(x)\to\int_{\reals}\phi(x)\d\mu(x)$ for all continuous functions $\phi:\reals\to\reals$ with at most quadratic growth (i.e., $\phi(x)\leq A(1+\abs{x}^2)$ for some $A>0$).

\section{Model}\label{sec:model}
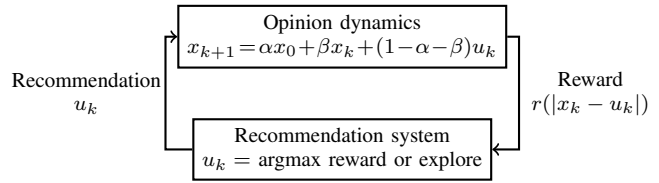
\begin{figure}[t]
    \centering
    \begin{tikzpicture}
        \node[rectangle,draw,thick,align=center] at (0,0) (o) {Opinion dynamics \\ $\opinion_{k+1}\!=\!\weightbias\opinion_0\!+\!\weightopinion \opinion_k\!+\!(1\!-\!\weightbias\!-\!\weightopinion)\rec_k$}; 
        \node[rectangle,draw,thick,align=center] at (0,-1.5) (r) {Recommendation system \\ $\rec_k=$ argmax reward or explore};

        \draw[->,thick] (o.east) -- (+2.35,0) |- (r.east);
        \draw[->,thick] (r.west) -| (-2.35,0) -- (o.west);

        \node[align=center] at (+3.3,-0.75) {Reward \\ $\reward(\abs{\opinion_k-\rec_k})$};
        \node[align=center] at (-3.4,-0.75) {Recommendation \\ $\rec_k$};
    \end{tikzpicture}
    \caption{Closed loop between a user (whose opinion dynamics are detailed in~\cref{subsec:user}) and a recommendation system (whose algorithm is detailed~\cref{subsec:reccomendation}).}
    \label{fig:closed loop}
\end{figure}

In this section, we present our model. It consists of two interconnected parts: the user model and the recommendation system model; see~\cref{fig:closed loop}.
We start with a detailed description of each component. Then, we illustrate the behavior of our model in a numerical example. 
Our model is based on the mathematical framework from~\cite{rossi2021closed}.

\subsection{Modeling of the users' opinion dynamics}\label{subsec:user}
We consider a large homogeneous population of users. The opinion of each user evolves according to the Friedkin-Johnson model~\cite{friedkin1990social}
\begin{equation}\label{eq:opinion dynamics}
    \opinion_{k+1}
    =
    \weightbias \opinion_0
    +
    \weightopinion \opinion_k + (1-\weightbias-\weightopinion) \rec_k,
\end{equation}
where $\opinion_k\in\reals$ is the user's opinion at time $k$, $\opinion_0$ is her/his opinion bias and initial opinion, and $\rec_k\in\reals$ is the recommendation at time $k$.  The parameters $\weightbias\in[0,1]$ and $\beta\in[0,1)$ (with $\weightbias+\weightopinion\leq 1$) arbitrate between the impact of the user's bias, the current opinion, and the received recommendation on the future opinion. All users in the population share the same parameters $\weightbias$ and $\weightopinion$ but have different biases. 
In particular, the bias/initial opinion distribution of the population is $\dopinion_0\in\Pp{2}{\reals}$. Similarly, we denote by $\dopinion_k$ the opinion distribution at time $k$ or, equivalently, the probability distribution associated with the opinion of a generic user.
Given the recommendation, the users produce a reward according to monotonically decreasing function $\reward:\nonnegativereals\to\nonnegativereals$, so that the reward at time $k$ is $\reward(\abs{\opinion_k-\rec_k})$. Intuitively, the closer the recommendation is to the user's opinion, the more the user appreciates the content and thus the higher the benefit for the recommendation system (e.g., more clicks, more time on the platform). The reward is the only observable quantity; in particular, users' opinions are private and not revealed to the recommendation system.

\subsection{Modeling of the recommendation system}\label{subsec:reccomendation}
The recommendation system aims at maximizing the reward. To do so, it outputs the recommendation that has generated the largest reward throughout the entire past until time $k$, with the exception of exploration steps, happening at the beginning of the horizon ($k=0$) and every $T$ steps ($k=nT$ for $n\in\naturals$). Thus, for $k\notin\{0,T,2T,\ldots\}$
\begin{equation*}
    \rec_k
    =
    \argmax_{u_0,\ldots,u_{k-1}} \{\reward(\abs{\opinion_0-\rec_0}),\ldots,\reward(\abs{\opinion_{k-1}-\rec_{k-1}})\}.
\end{equation*}
When exploring, the recommendation system samples a recommendation from the recommendation distribution $\drec\in\Pp{2}{\reals}$; i.e., $\rec_k\sim\drec$ for $k\in\{0,T,2T,\ldots\}$. This strategy is in line with the classical $\varepsilon$-greedy action selection in multi-armed bandit problems in reinforcement learning, which also outputs the reward-maximizing action but explores with some probability $\varepsilon$ (instead of at fixed time steps)~\cite[§2]{sutton2018reinforcement}.

\subsection{Discussion of the model}
A few comments are in order. 
First, we consider a homogeneous population, whereby all users have the same $\weightbias$ and $\weightopinion$. This way, we can study the effect of the recommendation systems on \emph{ensembles} of users, rather than on a specific user with given $\weightbias$ and $\weightopinion$, bias/initial opinion $\opinion_0$, and realizations of the random recommendations. When presenting our results in~\cref{sec:results}, we consider various populations to illustrate the roles of $\weightbias$ and $\weightopinion$.
Second, without loss of generality, we assume $\weightopinion\neq 1$. If $\weightopinion=1$ (i.e., users are infinitely stubborn), we trivially conclude $\opinion_{k+1}=\opinion_k$ and $\dopinion_k=\dopinion_0$ for all $k$.
Third, we do not include peer-to-peer interactions within the population, in line with our focus being the interaction between a recommendation system and its users. In any case, our model and analysis can be extended to include influences between the users (e.g., via graphons~\cite{lovasz2012large}). Nonetheless, we argue that, if interactions are to be considered, they do not happen \emph{between} the users, but rather \emph{through} the recommendation system. An example is collaborative filtering, where the system generates recommendations for a given user based on other users considered similar~\cite{su2009survey}.
Fourth, there are of course many other models of recommendation systems based on more sophisticated algorithms, which include more general exploration strategies, collaborative filtering, priors on the users, machine learning techniques, etc. We leave their study to future research. 
Fifth, as we shall see below, the reward function $\reward$ does not impact our analysis, as long as a mild (and, arguably, realistic) monotonicity assumption is satisfied.
Finally, to ease the notation, we assume that $\weightbias$, $\weightopinion$, and the recommendation distribution $\drec$ are time-invariant. Provided this time-varying behavior is sufficiently regular, our model and analysis extend to the time-varying setting. 

\subsection{Illustrative example}\label{subsec:model:simulation}
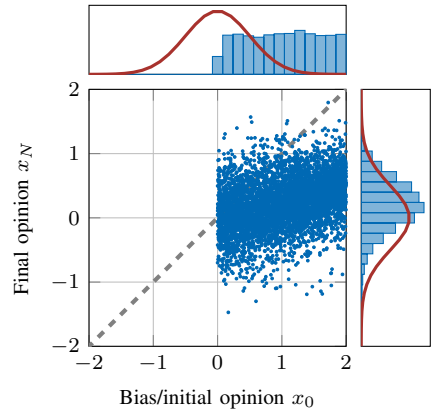
\begin{figure}[t]
    \centering
    \begin{tikzpicture}
        \begin{groupplot}[group style={group size=2 by 2, horizontal sep=0.2cm, vertical sep=0.2cm},
                          width=2.5cm, height=2.5cm]
            
            \nextgroupplot[width=5cm,
                           xmin=-2.0, xmax=2.0,
                           ymin=0, 
                           xtick=\empty, ytick=\empty]
            \addplot[hist={density,bins=25},ETHblue,fill=ETHblue,fill opacity=0.5] table [y expr=\thisrowno{0}, col sep=comma] {figures/illustrative_example.csv};
            \addplot[ETHred,very thick,domain=-2:2,samples=30] {1/(0.5*sqrt(2*3.14))*e^(-0.5*x^2/0.5^2)};

            \nextgroupplot[group/empty plot]

            \nextgroupplot[width=5cm, height=5cm,
                           grid=major,
                           xmin=-2.0, xmax=2.0, ymin=-2.0, ymax=2.0,
                           xlabel={Bias/initial opinion $\opinion_0$}, ylabel={Final opinion $\opinion_N$}]
            \addplot[ETHblue,only marks,mark size=0.5pt] table[x expr=\thisrowno{0}, y expr=\thisrowno{1}, col sep=comma] {figures/illustrative_example.csv};
            \addplot[gray, dashed, ultra thick] {x}; 

            \nextgroupplot[width=5cm,
                           xmin=-2.0, xmax=2.0,
                           ymin=0, 
                           ytick=\empty,xtick=\empty,rotate=-90,x dir=reverse]
            \addplot[hist={density,bins=25},ETHblue,fill=ETHblue,fill opacity=0.5] table [y expr=\thisrowno{1}, col sep=comma] {figures/illustrative_example.csv};
            \addplot[ETHred,very thick,domain=-2:2,samples=30] {1/(0.5*sqrt(2*3.14))*e^(-0.5*x^2/0.5^2)};
            
        \end{groupplot}
    \end{tikzpicture}
    \caption{Simulation of the closed-loop system over a horizon of 50 time steps. The central plot shows the final opinion of each user, as a function of their initial opinion. The top histogram shows the bias/initial opinion distribution; the one on the right is the final opinion distribution. For reference, we include, in solid red, the recommendation distribution.}
    \label{fig:simulations}
\end{figure}

To illustrate our model, we consider a population with $\weightbias=0.1$ and $\weightopinion=0.7$. For simulation purposes, we consider 5000 users, whose bias/initial opinion $\opinion_0$ is distributed uniformly on $[0,2]$ (blue, $x$-axis in~\cref{fig:simulations}). 
The recommendation distribution $\drec$ is a zero-mean Gaussian with a standard deviation of $0.5$ (red in~\cref{fig:simulations}) and exploration happens every $5$th time step. We run the system for $50$ time steps. 
Our results are in~\cref{fig:simulations}.
At a macroscopic level, the opinion distribution shifts towards the recommendation distribution so that the final opinion distribution $\dopinion_N$ is a slightly asymmetric Gaussian (right plot in~\cref{fig:simulations}). 
At a microscopic level, the opinion of most (but not all) users is lower than their initial opinion (central plot in~\cref{fig:simulations}), which collectively contributes to the centralization of the opinion distribution observed at a macroscopic level.

\section{Analysis}\label{sec:analysis}

In this section, we investigate some of the theoretic properties of our model. 
We start with a formulation of the dynamics in a tree structure, which immediately unveils the combinatorial nature of the problem. Thereafter, we study the more tractable yet insightful limit cases of $\exploration=+\infty$ (no exploration) and $\exploration=1$ (exploration at every time step).  

\subsection{The general case: A combinatorial problem}

A general analysis of the closed-loop system is combinatorial. After each exploration step, there are two possibilities: (i) the exploration input led to a higher reward and thus the recommendation system sticks to that recommendation for the subsequent steps, at least until the next exploration phase; (ii) the exploration input led to a lower reward and is therefore discarded, and the recommendation system ``goes back'' to the last recommendation. Accordingly, the probability of successful exploration coincides with the probability of sampling a recommendation (strictly) closer to the current opinion compared to the current recommendation. Formally:

\begin{lemma}[Probability of successful exploration]\label{lemma:exploration successful}
Let $\rec_k$ be the current recommendation and $\opinion_k$ be the opinion at time $k$. Then, the probability of successful exploration is
\begin{equation*}
    p(\opinion_k,\rec_k)
    =
    \begin{aligned}[t]
    &F_{\drec}(\opinion_k+|\opinion_k-\rec_k|)-\rho(\{\opinion_k+|\opinion_k-\rec_k|\})
    \\&-F_{\drec}(\opinion_k-|\opinion_k-\rec_k|),
    \end{aligned}
\end{equation*}
where $F_{\drec}:\reals\to[0,1]$ is the cumulative distribution function of the recommendation distribution $\drec$, and $\rho(\{a\})$ is the probability of sampling the recommendation $a\in\reals$.
\end{lemma}

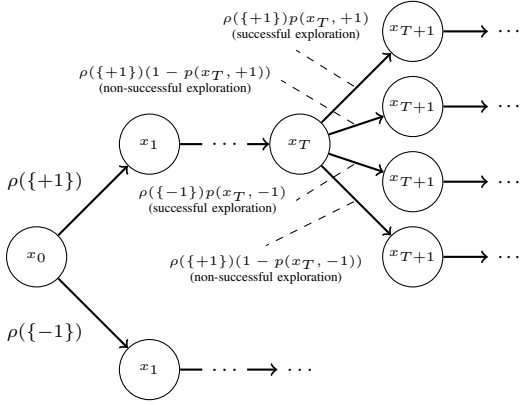
\begin{figure}[!t]
    \centering
    \tikzstyle{mycircle}=[circle,draw,font=\tiny,inner sep=1pt,minimum size=0.8cm]
    \tikzstyle{myarrow} = [thick]
    \begin{tikzpicture}
        \node[mycircle] at (-0.5,1.5) (o) {$\opinion_0$}; 

        \node[mycircle] at (1,3) (up1) {$\opinion_1$}; 
        \draw[->,myarrow] (o) -- (up1) node[pos=.5,above left,font=\scriptsize] {$\rho(\{+1\})$}; 
        
        \node at (2,3) (up2) {$\ldots$}; 
        \node[mycircle] at (3,3) (up3) {$\opinion_T$}; 
        \draw[-,myarrow] (up1) -- (up2);
        \draw[->,myarrow] (up2) -- (up3);
        
        \node[mycircle] at (4.5,4.5) (upup1) {$\opinion_{T+1}$}; 
        \node[mycircle] at (4.5,3.5) (upup2) {$\opinion_{T+1}$}; 
        \node[mycircle] at (4.5,2.5) (upup3) {$\opinion_{T+1}$};
        \node[mycircle] at (4.5,1.5) (upup4) {$\opinion_{T+1}$};
        \draw[->,myarrow] (up3) -- (upup1) node[pos=.5,font=\tiny,inner sep=0pt] (a1) {}; 
        \draw[->,myarrow] (up3) -- (upup2) node[pos=.5,font=\tiny,inner sep=0pt] (a2) {};
        \draw[->,myarrow] (up3) -- (upup3) node[pos=.5,font=\tiny,inner sep=0pt] (a3) {}; 
        \draw[->,myarrow] (up3) -- (upup4) node[pos=.5,font=\tiny,inner sep=0pt] (a4) {};
        \draw[->,myarrow] (upup1) --++ (1,0) node[pos=1,right] (b1) {\ldots}; 
        \draw[->,myarrow] (upup2) --++ (1,0) node[pos=1,right] (b2) {\ldots};
        \draw[->,myarrow] (upup3) --++ (1,0) node[pos=1,right] (b3) {\ldots}; 
        \draw[->,myarrow] (upup4) --++ (1,0) node[pos=1,right] (b4) {\ldots};

        \draw[-,dashed] (a1) -- ($(a1)+(-0.8,0.5)$) node[pos=1,above,font=\tiny,align=center] {$\rho(\{+1\})p(\opinion_T,+1)$ \\ (successful exploration)}; 
        \draw[-,dashed] (a2) -- ($(a2)+(-1,0.6)$) node[pos=1,left,font=\tiny,align=center] {$\rho(\{+1\})(1-p(\opinion_T,+1))$ \\ (non-successful exploration)}; 
        \draw[-,dashed] (a3) -- ($(a3)+(-0.8,-0.5)$) node[pos=1,left,font=\tiny,align=center] {$\rho(\{-1\})p(\opinion_T,-1)$\\ (successful exploration)}; 
        \draw[-,dashed] (a4) -- ($(a4)+(-1.2,-0.6)$) node[pos=1,below,font=\tiny,align=center] {$\rho(\{+1\})(1-p(\opinion_T,-1))$\\ (non-successful exploration)};

        \node[mycircle] at (1,0) (down1) {$\opinion_1$}; 
        \draw[->,myarrow] (o) -- (down1) node[pos=.5,below left,font=\scriptsize] {$\rho(\{-1\})$}; 
        \node at (2,0) (down2) {$\ldots$}; 
        \node at (3,0) (down3) {$\ldots$}; 
        \draw[-,myarrow] (down1) -- (down2);
        \draw[->,myarrow] (down2) -- (down3);
    \end{tikzpicture}
    \caption{Dynamics in the simple case of two possible recommendations ($\pm 1$, i.e., $\drec=p \delta_{-1}+(1-p)\delta_{+1}$ for $p\in[0,1]$) and deterministic bias/initial opinion $\opinion_0\in\reals$ (i.e., $\dopinion_0=\delta_{\opinion_0}$). The transitions are $\opinion_{k+1}=\weightbias\opinion_0 +\weightopinion\opinion_k+(1-\weightbias-\weightopinion)\rec_k$; the quantity associated with each arrow is the probability of that transition. 
    }
    \label{fig:tree}
\end{figure}

\cref{lemma:exploration successful} predicates that the probability of successful exploration is controlled by the difference between the current opinion and recommendation, together with the properties of the recommendation distribution $\rho$. For instance, if $\drec$ is uniform between $-1$ and $+1$, then $\rho(\{x\})=0$ for all $x\in\reals$ and
$F_\rho(x)=\frac{1}{2}(1+\max\{-1,\min\{1,x\}\})$, so that 
$p(\opinion_k,\rec_k)=\frac{1}{2}(1+\max\{-1,\min\{1, \opinion_k+\abs{\opinion_k-\rec_k}\}\})-\frac{1}{2}(1+\max\{-1,\min\{1,\opinion_k-\abs{\opinion_k-\rec_k}\}\})$.
If $\opinion_k\pm\abs{\opinion_k-\rec_k}\in[-1,+1]$, then $p(\opinion_k,\rec_k)=\abs{\opinion_k-\rec_k}$, showing that, at least in the case of a uniform recommendation distribution, the probability of successful exploration is precisely $\abs{\opinion_k-\rec_k}$.
With~\cref{lemma:exploration successful}, the dynamics are as in~\cref{fig:tree}, which explains why the analysis of the closed-loop system becomes intractable after a few exploration steps. Thus, in the sequel, we restrict our analysis to two tractable yet insightful limit cases: exploration only at the initial time and exploration at every time step. 

\subsection{Special case: No exploration}
Suppose now that the recommendation system does not perform exploration; i.e., $\exploration\to+\infty$. In this case, the initial recommendation, which is random, will be applied at all time steps, regardless of the reward returned by the user. As a result, each user's opinion converges to a convex combination of the bias and the received recommendation (at the initial time). Macroscopically, the opinion distribution approaches a ``convex combination'' of the bias distribution and the recommendation distribution:

\begin{proposition}[No exploration]
\label{prop:specialcase:noexploration}
Let $T=+\infty$.
Let $\dopinion_0\in\Pp{2}{\reals}$ be the bias/initial opinion distribution, $\dopinion_k\in\Pp{2}{\reals}$ be the opinion distribution at time $k$, and $\drec\in\Pp{2}{\reals}$ be the recommendation distribution.
Then, the opinion distribution $\dopinion_k$ converges weakly in $\Pp{2}{\reals}$ to the opinion distribution
\begin{equation}\label{eq:specialcase:exploration:limit}
    \mu\coloneqq 
    \pushforward{\left(\frac{\weightbias}{1-\weightopinion}x\right)}\dopinion_0
    \ast 
    \pushforward{\left(\left(1-\frac{\weightbias}{1-\weightopinion}\right)x\right)}\drec.
\end{equation}
\end{proposition}

\cref{prop:specialcase:noexploration} proves weak convergence of the opinion distribution (i.e., convergence of the integral of each continuous function which grows at most quadratically) and does not prove strong convergence (i.e., $\mu_k(A)\to\mu(A)$ for every Borel set $A$). Namely, we prove that all macroscopic quantities converge (e.g., $\phi(x)=x$ in the definition of weak convergence in $\Pp{2}{\reals}$ yields convergence of the expected value), and refrain from conducting a microscopic analysis for each and every (infinitesimal) user of the population. In particular, if $\weightbias=0$, then the opinion distribution asymptotically converges to the recommendation distribution. 
While simple,~\cref{prop:specialcase:noexploration} unveils a fundamental discrepancy between a \emph{macroscopic} analysis, aiming to study the opinion distribution of the population, and a \emph{microscopic} analysis, which considers the change of opinions for individual users. We illustrate the phenomenon in the following analytic example and discuss it more in detail when presenting our numerical results in~\cref{sec:results}. 

\begin{example}[Microscopic vs. macroscopic behavior]\label{example:micro vs macro}
Suppose that the bias/initial opinion distribution $\dopinion_0$ coincides with the recommendation distribution $\drec$ and that all distributions are zero-mean Gaussian with standard deviation $\sigma>0$. By \cref{prop:specialcase:noexploration} (together with the expressions for the pushforward and convolution of Gaussians), the opinion distribution converges to a zero-mean Gaussian with standard deviation $\sigma\sqrt{(\frac{\weightbias}{1-\weightopinion})^2+(1-\frac{\weightbias}{1-\weightopinion})^2}$, so that the (type-2) Wasserstein distance between the bias/initial opinion distribution $\dopinion_0$ and the final distribution $\dopinion$ (defined in~\eqref{eq:specialcase:exploration:limit}) reads
\begin{equation*}
    \wasserstein{2}{\dopinion_0}{\mu}^2
    =
    \sigma^2\abs{1-\sqrt{\left(\frac{\weightbias}{1-\weightopinion}\right)^2+\left(1-\frac{\weightbias}{1-\weightopinion}\right)^2}}^2,
\end{equation*}
where we used the closed-form expression for the Wasserstein distance between Gaussians.
Conversely, from a microscopic perspective, the opinion of a user with bias $\opinion_0$ who receives the recommendation $\rec_0$ converges to $\frac{\weightbias}{1-\weightopinion}\opinion_0+(1-\frac{\weightbias}{1-\weightopinion})\rec_0$. Thus, the opinion shift is $(1-\frac{\weightbias}{1-\weightopinion})(\opinion_0-\rec_0)$, and the expected squared opinion shift for a user is
\begin{equation*}
\begin{aligned}
    \Delta
    &=
    \abs{1-\frac{\weightbias}{1-\weightopinion}}^2\int_{\reals}\int_{\reals}\abs{\opinion_0-\rec_0}^2\d\dopinion_0(\opinion_0)\d\drec(\rec_0)
    \\
    &=
    2\abs{1-\frac{\weightbias}{1-\weightopinion}}^2\sigma^2.
\end{aligned}
\end{equation*}
For $\weightbias\to 1-\weightopinion$, we recover the trivial case $1-\weightbias-\weightopinion=0$, whereby the recommendation has weight 0 in the opinion dynamics~\eqref{eq:opinion dynamics}. Accordingly, both $\wasserstein{2}{\dopinion_0}{\mu}^2$ and $\Delta$ converge to 0, and the micro- and macroscopic behaviors align.
For $\weightbias\to 0$ and fixed $\weightopinion$, instead, $\wasserstein{2}{\dopinion_0}{\mu}^2\to 0$; i.e., the initial opinion distribution $\dopinion_0$ (which, by assumption, equals $\drec$) coincides with the final opinion distribution $\mu$ (which, by~\cref{prop:specialcase:noexploration}, also equals  $\drec$). However, $\Delta\to 2\sigma^2>0$. Thus, microscopically, each user's opinion is highly impacted by the recommendation system, while, macroscopically, the opinion distribution of the population is unaltered. 
\end{example}

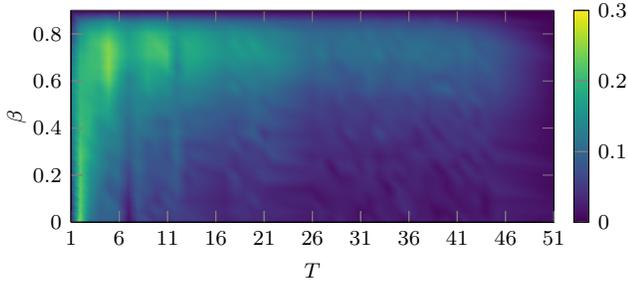
\begin{figure}[t]
    \centering 
    \begin{tikzpicture}
    \begin{axis}[name=one,
                 colorbar, colorbar style={width=0.2cm,xshift=-0.3cm},
                 width=8.0cm, height=4.4cm,
                 view={0}{90},
                 point meta min=0, point meta max=0.3,
                 xlabel={$\exploration$},
                 xmin=1, xmax=51,
                 xtick={1,6,...,50},
                 ylabel={$\weightopinion$}, ylabel style={yshift=-0.15cm},
                 ymin=0, ymax=0.9,
                 ytick={0,0.2,0.4,0.6,0.8},
                 shader=interp,
                 colormap/viridis,
                 ]
    \addplot3[surf,mesh/cols=20] table [x expr=\thisrowno{0}, y expr=\thisrowno{1}, z expr=\thisrowno{2}, col sep=comma] {figures/wasserstein_distance_convergence_no_exploration.csv}; 
    \end{axis}
    \end{tikzpicture}
    \caption{The (type-1) Wasserstein distance between the final opinion distribution and the distribution~\eqref{eq:specialcase:exploration:limit} is small (dark blue in the colormap) already for relatively small values $T$, especially for $\weightopinion$ small (i.e., the distance is small). Thus, at least qualitatively,~\cref{prop:specialcase:noexploration} remains valid also in non-asymptotic regimes. For this simulation, we used the same setting as in~\cref{subsec:model:simulation}.}
    \label{fig:noexploration}
\end{figure}

Finally, \cref{prop:specialcase:noexploration} considers the limit setting where the recommendation system explores only at the beginning and then sticks to the first, random, recommendation. Nonetheless, the intuition remains valid for sufficiently large values of exploration, as suggested by the numerical results in~\cref{fig:noexploration}.

\subsection{Special case: Continuous exploration}

We consider now the opposite limit case, where the recommendation system explores at each time step; i.e., $T=1$ and the recommendation received by a user is therefore random at each time step. In this setting, the opinion distribution converges to a ``convex combination'' of the bias/initial opinion distribution and a distribution ``similar'' to a Gaussian distribution. Perhaps surprisingly, this result is independent of the initial opinion distribution and, importantly, of the recommendation distribution. Formally: 

\begin{proposition}[Continuous exploration]
\label{prop:specialcase:continuous exploration}
Let $T=1$.
Let $\dopinion_0\in\Pp{2}{\reals}$ be the bias/initial opinion distribution, $\dopinion_k\in\Pp{2}{\reals}$ be the opinion distribution at time $k$, and $\drec\in\Pp{2}{\reals}$ be the recommendation distribution.
Then, the opinion distribution $\dopinion_k$ converges weakly in $\Pp{2}{\reals}$ to some opinion distribution $\dopinion\in\Pp{2}{\reals}$ with
\begin{equation}\label{eq:always exploration:mean and variance}
\begin{aligned}
    \expectedValue{\dopinion}{x}
    &=
    \frac{\weightbias}{1-\weightopinion}\expectedValue{\dopinion_0}{x}
    +\left(1-\frac{\weightbias}{1-\weightopinion}\right)
    \expectedValue{\drec}{x}
    \\
    \variance(\dopinion)
    &=
    \frac{\weightbias^2}{(1-\weightopinion)^2}\variance(\dopinion_0)
    +
    \frac{(1-\weightbias-\weightopinion)^2}{1-\weightopinion^2}\variance(\drec).
\end{aligned}
\end{equation}
Moreover, regardless of the bias/initial opinion distribution $\dopinion_0$ and the recommendation distribution $\drec$,
\begin{equation}\label{eq:always exploration:convergence}
    \dopinion = \pushforward{\left(\frac{\weightbias}{1-\weightopinion}x\right)}\dopinion_0\ast
    \pushforward{\left(\left(1-\frac{\weightbias}{1-\weightopinion}\right)x\right)}\bar\drec,
\end{equation}
where $\bar\rho$  is ``almost Gaussian'', in the sense that its normalized distribution $\hat\drec\coloneqq\pushforward{((x-\expectedValue{\bar\drec}{x})/\sqrt{\variance(\bar\drec)})}\bar\drec$  satisfies
\begin{equation}\label{eq:always exploration:almost gaussian}
    \wasserstein{1}{\hat\drec}{\Phi}
    \leq 
    \left(\frac{18}{\pi}\right)^{\frac{1}{3}}\left(\frac{1-\weightopinion^2}{e\weightopinion^2}\right)^{\frac{1}{12}}\sup_{\xi\neq 0}\frac{\abs{C_{\drec}(\xi)-C_{\Phi}(\xi)}}{\abs{\xi}^3},
\end{equation}
where $\Phi$ is the zero-mean Gaussian probability distribution with unit variance and $C_{\dopinion}(\cdot)$ is the characteristic function of $\dopinion$.
Finally, if $\dopinion_0$ and $\drec$ are Gaussian distributions, then $\dopinion$ is Gaussian with the mean and variance in~\eqref{eq:always exploration:mean and variance}.
\end{proposition}

\begin{figure}[bt]
    \centering
    \begin{tikzpicture}
        \begin{groupplot}[group style={group size=2 by 2, horizontal sep=0.2cm, vertical sep=0.2cm},
                          width=2.5cm, height=2.5cm]
            
            \nextgroupplot[width=5cm,
                           xmin=-2.0, xmax=2.0,
                           ymin=0, 
                           xtick=\empty, ytick=\empty]
            \addplot[hist={density,bins=25},ETHblue,fill=ETHblue,fill opacity=0.5] table [y expr=\thisrowno{0}, col sep=comma] {figures/gaussian.csv};
            \addplot[ETHred,very thick,domain=-2:2,samples=30] {0.5/(0.4*sqrt(2*3.14))*(e^(-0.5*(x-1)^2/0.4^2)+e^(-0.5*(x+1)^2/0.4^2))};

            \nextgroupplot[group/empty plot]

            \nextgroupplot[width=5cm, height=5cm,
                           grid=major,
                           xmin=-2.0, xmax=2.0, ymin=-2.0, ymax=2.0,
                           xlabel={Bias/initial opinion $\opinion_0$}, ylabel={Final opinion $\opinion_N$}]
            \addplot[ETHblue,only marks,mark size=0.5pt] table[x expr=\thisrowno{0}, y expr=\thisrowno{1}, col sep=comma] {figures/gaussian.csv};
            \addplot[gray, dashed, ultra thick] {x}; 

            \nextgroupplot[width=5cm,
                           xmin=-2.0, xmax=2.0,
                           ymin=0, 
                           ytick=\empty,xtick=\empty,rotate=-90,x dir=reverse]
            \addplot[hist={density,bins=25},ETHblue,fill=ETHblue,fill opacity=0.5] table [y expr=\thisrowno{1}, col sep=comma] {figures/gaussian.csv};
            \addplot[ETHred,very thick,domain=-2:2,samples=30] {0.5/(0.4*sqrt(2*3.14))*(e^(-0.5*(x-1)^2/0.4^2)+e^(-0.5*(x+1)^2/0.4^2))};
            
        \end{groupplot}
    \end{tikzpicture}
    \caption{Simulation of the closed-loop system over a horizon of 100. As suggested by our theoretic results in the case of infinitely frequent exploration (and with $\weightbias=0$), the opinion distribution approaches (but does not generally converge to) a Gaussian distribution, even though the bias/initial opinion distribution is uniform and the recommendation distribution (plotted in solid red) is bimodal (a mixture of Gaussians with mean $\pm 1$ and standard deviation 0.4).}
    \label{fig:gaussian}

\end{figure}
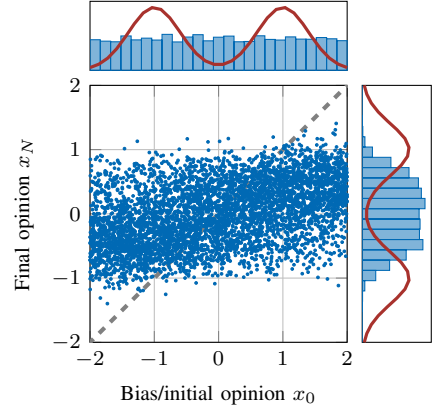

\cref{prop:specialcase:continuous exploration} predicates that the opinion distribution converges to a ``convex combination'' of the bias/initial condition distribution and a distribution which is ``almost Gaussian'', regardless of the recommendation distribution (that is, even for very ``non-Gaussian'' recommendation distributions).
This effect is reminiscent of the central limit theorem in probability theory, which, however, does not apply here (indeed, we consider trajectories of a stochastic dynamic system and not the sum of i.i.d. random variables). In particular, if $\weightbias=0$, the opinion distribution will asymptotically be ``almost Gaussian''.
The upper bound~\eqref{eq:always exploration:almost gaussian}, which is valid for all $\weightbias$ and $\weightopinion$ and is non-trivial (i.e., finite) for all distributions with finite third moment~\cite{katriel2019quantitative}, indicates that the vicinity to a Gaussian distribution increases continuously as $\weightopinion$ approaches 1. When all distributions are Gaussian, the result becomes exact. 
Among others, \cref{prop:specialcase:continuous exploration} explains why, for short exploration times, the final opinion distribution does not depend on the recommendation distribution and why, for small values of $\weightbias$, it resembles a Gaussian distribution, even if all underlying distributions are not Gaussian; see~\cref{fig:gaussian}, where $\weightbias=0$, $\weightopinion=0.8$, and $\exploration=3$.

\section{Numerical Results}\label{sec:results}

Our numerical analysis\footnote{Python implementation: \url{https://gitlab.ethz.ch/lnicolas/impact-of-recommendation-systems-on-opinion-dynamics}.}
concerns the discrepancy between the micro- and macroscopic behavior of the opinion distribution, as a function of $\weightbias, \weightopinion$ and the exploration time $\exploration$.
We consider homogeneous populations with $\weightbias\in\{0,0.1,0.2\}$, $\weightopinion\in\{0,0.1,\ldots,1-\weightbias\}$, and bias/initial opinion $\opinion_0$ uniformly distributed between $-2$ and $2$, and recommendation systems with exploration time $T\in\{1,\ldots,21\}$ and recommendations distributed according to the standard zero-mean Gaussian distribution with unit variance. For each setting, we perform 20 exploration cycles. 
To quantify \emph{microscopic} opinion shifts, we average the difference between users' initial and final opinions (i.e., $\frac{1}{M}\sum_{i=1}^M\abs{\opinion_{0,i}-\opinion_{N,i}}$ with $\opinion_{0,i}$ and $\opinion_{N,i}$ being the initial and final opinion of user $i\in\{1,\ldots,M\}$).
To quantify \emph{macroscopic} opinion shifts, instead, we use the (type-1) Wasserstein distance between bias/initial opinion distribution $\dopinion_0$ and final opinion distribution $\dopinion_N$.

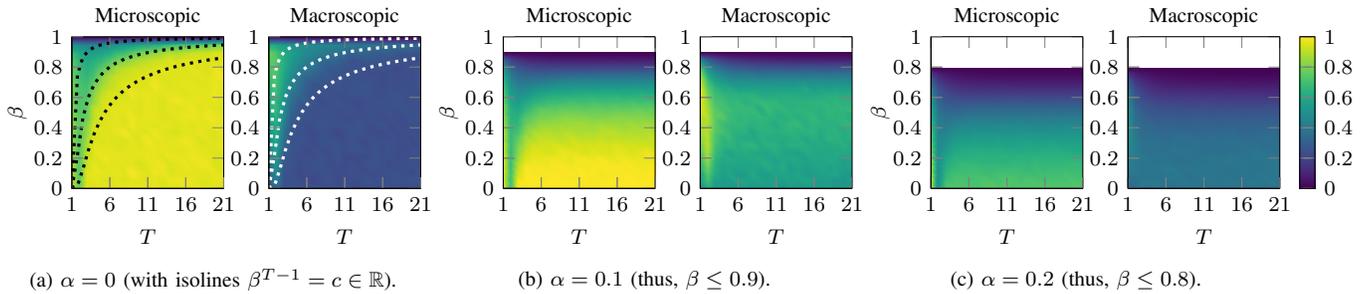
\begin{figure*}[bt]
    \centering
    
    \hspace{-0.9cm}
    \begin{subfigure}[t]{0.32\textwidth}
    \begin{tikzpicture}
    \begin{groupplot}[name=one,
                      width=3.6cm, height=3.6cm,
                      view={0}{90},
                      point meta min=0, point meta max=1.0,
                      xlabel={$\exploration$},
                      xmin=1, xmax=21,
                      xtick={1,6,...,19},
                      ylabel={$\weightopinion$}, ylabel style={yshift=-0.15cm},
                      ymin=0, ymax=1,
                      ytick={0,0.2,0.4,0.6,0.8,1},
                      shader=interp,
                      colormap/viridis,
                      title style={yshift=-0.2cm},
                      group style = {group size=2 by 1,
                                     horizontal sep=0.6cm},
                ]
    \nextgroupplot[title={Microscopic}]
    \addplot3[surf,mesh/cols=20] table [x expr=\thisrowno{0}, y expr=\thisrowno{1}, z expr=\thisrowno{2}, col sep=comma] {figures/d_micro_alpha=0.0.csv}; 
    \addplot[samples=30,domain=1.01:22,very thick, dotted, black] {0.81^(1/(x-1))};
    \addplot[samples=30,domain=1.01:22,very thick, dotted, black] {0.33^(1/(x-1))};
    \addplot[samples=30,domain=1.01:22,very thick, dotted, black] {0.05^(1/(x-1))};

    \nextgroupplot[title={Macroscopic},ylabel={}]
    \addplot3[surf,mesh/cols=20] table [x expr=\thisrowno{0}, y expr=\thisrowno{1}, z expr=\thisrowno{2}, col sep=comma] {figures/d_macro_alpha=0.0.csv}; 
    \addplot[samples=30,domain=1.01:22,very thick, dotted, white] {0.81^(1/(x-1))};
    \addplot[samples=30,domain=1.01:22,very thick, dotted, white] {0.33^(1/(x-1))};
    \addplot[samples=30,domain=1.01:22,very thick, dotted, white] {0.05^(1/(x-1))};
    \end{groupplot}
    \end{tikzpicture}
    \caption{$\weightbias=0$ (with isolines $\weightopinion^{\exploration-1}=c\in\reals$).}
    \label{fig:results:micro macro:alpha 0}
    \end{subfigure}
    \hspace{-0.2cm}
    \begin{subfigure}[t]{0.32\textwidth}
    \begin{tikzpicture}
    \begin{groupplot}[name=one,
                      width=3.6cm, height=3.6cm,
                      view={0}{90},
                      point meta min=0, point meta max=1.0,
                      xlabel={$\exploration$},
                      xmin=1, xmax=21,
                      xtick={1,6,...,19},
                      ylabel={$\weightopinion$}, ylabel style={yshift=-0.15cm},
                      ymin=0, ymax=1,
                      ytick={0,0.2,0.4,0.6,0.8,1},
                      shader=interp,
                      colormap/viridis,
                      title style={yshift=-0.2cm},
                      group style = {group size=2 by 1,
                                     horizontal sep=0.6cm},
                ]
    \nextgroupplot[title={Microscopic}]
    \addplot3[surf,mesh/cols=20] table [x expr=\thisrowno{0}, y expr=\thisrowno{1}, z expr=\thisrowno{2}, col sep=comma] {figures/d_micro_alpha=0.1.csv}; 

    \nextgroupplot[title={Macroscopic},ylabel={}]
    \addplot3[surf,mesh/cols=20] table [x expr=\thisrowno{0}, y expr=\thisrowno{1}, z expr=\thisrowno{2}, col sep=comma] {figures/d_macro_alpha=0.1.csv}; 
    \end{groupplot}
    \end{tikzpicture}
    \caption{$\weightbias=0.1$ (thus, $\weightopinion\leq 0.9$).}
    \end{subfigure}
    \hspace{-0.2cm}
    \begin{subfigure}[t]{0.32\textwidth}
    \begin{tikzpicture}
    \begin{groupplot}[name=one,
                      width=3.6cm, height=3.6cm,
                      view={0}{90},
                      point meta min=0, point meta max=1.0,
                      xlabel={$\exploration$},
                      xmin=1, xmax=21,
                      xtick={1,6,...,19},
                      ylabel={$\weightopinion$}, ylabel style={yshift=-0.2cm},
                      ymin=0, ymax=1,
                      ytick={0,0.2,0.4,0.6,0.8,1},
                      shader=interp,
                      colormap/viridis,
                      title style={yshift=-0.2cm},
                      group style = {group size=2 by 1,
                                     horizontal sep=0.6cm},
                ]
    \nextgroupplot[title={Microscopic}]
    \addplot3[surf,mesh/cols=20] table [x expr=\thisrowno{0}, y expr=\thisrowno{1}, z expr=\thisrowno{2}, col sep=comma] {figures/d_micro_alpha=0.2.csv}; 

    \nextgroupplot[title={Macroscopic},ylabel={},colorbar,colorbar style={width=0.2cm,xshift=-0.3cm}]
    \addplot3[surf,mesh/cols=20] table [x expr=\thisrowno{0}, y expr=\thisrowno{1}, z expr=\thisrowno{2}, col sep=comma] {figures/d_macro_alpha=0.2.csv}; 
    \end{groupplot}
    \end{tikzpicture}
    \caption{$\weightbias=0.2$ (thus, $\weightopinion\leq 0.8$).}
    \end{subfigure}

    \caption{
    In each figure, the left plot shows the microscopic opinion shift as a function of $\weightopinion$ and $\exploration$, measured as the average difference between the initial and the final opinion across the population. The right plot shows the macroscopic opinion shift, captured by the (type-1) Wasserstein distance between the initial opinion distribution and the final one. For comparison purposes, we normalize each value by the maximum value across all $\weightopinion$ and $\exploration$, so that all values are between $0$ and $1$.
    Remarkably, our result highlights a marked discrepancy between the micro- and macroscopic behavior: Even if the opinion distribution can be shown to remain stable (e.g., via surveys), the opinion of individual users might be significantly impacted by the recommendation system, especially for low $\weightbias$.}
    \label{fig:results:micro macro}
\end{figure*}

Our results are summarized in~\cref{fig:results:micro macro}.
First, for small values of $\weightbias$, our simulations suggest a qualitative discrepancy between micro- and macroscopic changes: When the microscopic change is largest (yellow), the macroscopic change is lowest (blue), and vice versa. The trend becomes less marked and eventually disappears as $\weightbias$ increases, as already observed in~\cref{example:micro vs macro} for the limit case $T=+\infty$. This discrepancy suggests that, even if the opinion distribution can be proven to be relatively stable (e.g., through surveys), the opinions of individual users may be significantly impacted by the recommendation system.  
Second, for $\weightbias\ll 1-\weightopinion$, we observe similar behavior along isolines of $\weightopinion^{\exploration-1}=c\in\nonnegativereals$ in the sense that settings with similar $\weightopinion^{\exploration-1}$ yield similar micro- and macroscopic behaviors (see~\cref{fig:results:micro macro:alpha 0}). Intuitively, this is a consequence of the probability of successful exploration being modulated by $\abs{\opinion_k-\rec_k}$ (cf.~\cref{lemma:exploration successful}), since between two exploration steps $k$ and $k-T$: 
\begin{equation*}
\begin{aligned}
    \abs{\opinion_{k}-\rec_{k}} 
    \overset{\eqref{eq:opinion dynamics}}&{=}
    \abs{\weightopinion^{T-1}(\opinion_{k+1-T}-\rec_k) + \frac{\weightbias-\weightbias\weightopinion^{T-1}}{1-\weightopinion}(\opinion_0-\rec_k)}
    \\
    \overset{\weightbias\ll 1-\weightopinion}&{\approx} \beta^{T-1}\abs{\opinion_{k+1-T}-\rec_k}.
\end{aligned}
\end{equation*}
Thus, settings with similar $\weightopinion^{\exploration-1}$ result in similar probabilities of successful exploration and thus in similar outcomes.

\section{Conclusions}\label{sec:conclusion}
We studied the impact of recommendation systems on opinion dynamics from a microscopic (i.e., at the individual level) and macroscopic (i.e., at the population level) perspective. We analyzed theoretically and numerically the interaction between users and a recommendation system.
Among others, our work explains why and in which circumstances we observe a fundamental discrepancy between micro- and macroscopic effects, whereby the opinions of individual users are drastically affected by the recommendation system whereas the opinion distribution of the population is unchanged.

In future research, we would like to (i) further investigate the properties of our model, (ii) consider more sophisticated recommendation systems (e.g., collaborative filtering~\cite{su2009survey}), and (iii) model the dynamics of opinion distribution directly in the probability space (e.g., see~\cite{lanzetti2022modeling}). 


\bibliographystyle{IEEEtran}
\bibliography{references.bib}


\ifbool{finalsubmission}{}{
\appendix
\subsection{Preliminaries}

We start with two properties of the Wasserstein distance:  

\begin{lemma}[Pushforward and $\wassersteinLetter{2}$~\protect{\cite[Proposition 3]{aolaritei2022uncertainty}}]\label{lemma:wasserstein pushforward}
Let $\mu,\nu\in\Pp{}{\reals}$ and $f,g:\reals\to\reals$ be Borel maps. Then, 
\begin{equation*}
    \wasserstein{2}{\pushforward{f}\mu}{\pushforward{g}\nu}
    =
    \sqrt{\inf_{\gamma\in\setPlans{\mu}{\nu}}\int_{\reals\times\reals}\abs{f(x)-g(y)}^2\d\gamma(x,y)}.
\end{equation*}
Moreover, $\wasserstein{2}{\pushforward{f}\mu}{\pushforward{g}\mu}\leq \sqrt{\int_{\reals}\abs{f(x)-g(x)}^2\d\mu(x)}$, and, if $f(x)=ax$, $\wasserstein{2}{\pushforward{f}\mu}{\pushforward{f}\nu}\leq  a\wasserstein{2}{\mu}{\nu}$.
\end{lemma}

\begin{lemma}[Convolution and $\wassersteinLetter{2}$~\protect{\cite[Proposition 10]{aolaritei2022uncertainty}}]\label{lemma:wasserstein convolution}
Let $\mu,\nu,\rho\in\Pp{}{\reals}$. Then, 
    $\wasserstein{2}{\mu\ast\rho}{\nu\ast\rho}
    \leq 
    \wasserstein{2}{\mu}{\nu}$.
\end{lemma}

Second, we study the discounted sum of i.i.d. random variables. Here, $\mathcal{B}([0,1])$ denotes the Borel sigma algebra on $[0,1]$ and $\lambda$ the Lebesgue measure. 

\begin{lemma}[``Central limit theorem'' for discounted sums]\label{lemma:discounted clt}
Consider i.i.d. real random variables $(X_n)_{n\in\naturals}$ with mean $m$, variance $\sigma^2$, and law $\nu\in\Pp{}{\reals}$ on a probability space $([0,1],\mathcal{B}([0,1]),\lambda)$.
Let  $\beta\in(0,1)$ and define the random variable $S_\beta$ as the almost sure limit of $S_{\beta,k}
    \coloneqq 
    \sum_{l=0}^{k-1}\beta^{k-1-l} (1-\beta) X_l$.
Let $\hat\mu_\beta\in\Pp{}{\reals}$ be the law of the ``normalized'' $S_\beta$ (i.e., $\hat S_\beta=\sqrt{\frac{1+\beta}{(1-\beta)\sigma^2}}(S_\beta-\mu)$). Then, 
\begin{equation*}
    \wasserstein{1}{\hat\mu_\beta}{\Phi}
    \leq 
    \left(\frac{18}{\pi}\right)^{\frac{1}{3}}\left(\frac{1-\beta^2}{e\beta^2}\right)^{\frac{1}{12}}\sup_{\xi\neq 0}\frac{\abs{C_{\nu}(\xi)-C_{\Phi}(\xi)}}{\abs{\xi}^3},
\end{equation*}
where $C_{\mu}(\cdot)$ is the characteristic function of $\mu$ and $\Phi$ is the zero-mean Gaussian probability measure with unit variance. 
\end{lemma}

\begin{proof}
Without loss of generality, assume $X_l$ is zero-mean; indeed, if $\sum_{l=0}^{k}\beta^{k-1-l}(1-\beta)(X_l-m)$ converges to $S_\beta$, then $\sum_{l=0}^{k}\beta^{k-1-l}(1-\beta)X_l$ converges to $S_\beta+m$. Then, $S_\beta$ is well-defined: Standard arguments of the proof of the Khintchine-Kolmogorov Convergence Theorem~\cite[Theorem 1, §5]{chow2003probability} yield almost sure convergence. 
The rest of the proof is analogous to the proof of~\cite{katriel2019quantitative} (with $\beta$ instead of $a$ and for simplicity $s=3$), together with~\cite[Theorem 2.21]{carrillo2007contractive} to derive the bound for the Wasserstein distance.
\end{proof}

\subsection{Proofs}

\begin{proof}[Proof of~\cref{lemma:exploration successful}]
By definition of successful exploration,
\begin{align*}
    p
    &=\drec(\{u\in\reals:\reward(\abs{\opinion_k-u})<\reward(\abs{\opinion_k-\rec_k}\}))
    \\
    \overset{\heartsuit}&{=}\drec(\{u\in\reals:\abs{\opinion_k-u}< \abs{\opinion_k-\rec_k}\})
    \\
    &=\drec(\{u\in\reals:\opinion_k-\abs{\opinion_k-\rec_k}<u<\opinion_k+\abs{\opinion_k-\rec_k}\})
    \\
    &=
    \begin{aligned}[t]
    &\drec(\{u\in\reals:u\leq \opinion_k+\abs{\opinion_k-\rec_k}\})
    \\ &-\drec(\{\opinion_k+\abs{\opinion_k-\rec_k}\})
    \\ &-\drec(\{u\in\reals:u\leq \opinion_k-\abs{\opinion_k-\rec_k}\}),
    \end{aligned}
\end{align*}
where $\heartsuit$ follows from strict monotonicity of the reward. With the definition of $F_\drec$, we conclude the proof. 
\end{proof}

\begin{proof}[Proof of~\cref{prop:specialcase:noexploration}]
To start, consider a user with bias/initial opinion $\opinion_0\in\reals$ and let $u_0\in\reals$ be the recommendation at the initial time. The opinion at time $k$ is
\begin{equation}\label{eq:no exploration:proof:user}
\begin{aligned}
    \opinion_{k}
    &=
    \weightopinion^k \opinion_0
    +
    \sum_{l=0}^{k-1}\weightopinion^{k-1-l}\left(\weightbias\opinion_0+(1-\weightbias-\weightopinion)\rec_0\right)
    \\
    &=
    \frac{\weightbias+\weightopinion^k(1-\weightbias-\weightopinion)}{1-\weightopinion}\opinion_0 \!+\!
    \frac{(1-\weightopinion^k)(1-\weightbias-\weightopinion)}{1-\weightopinion}u_0
    \\
    &=
    \left(\eta+\weightopinion^k(1-\eta)\right)\opinion_0+
    (1-\weightopinion^k)(1-\eta) u_0,
\end{aligned}
\end{equation}
where $\eta\coloneqq\weightbias/(1-\weightopinion)\in[0,1]$. Without loss of generality, assume $\eta\neq 1$; else $\weightbias=1-\weightopinion$, and we directly conclude $\opinion_k=\opinion_0$ and so  $\dopinion_k=\dopinion_0$ for all $k\in\naturals$.
By~\eqref{eq:no exploration:proof:user}, the opinion distribution at time $k$ is 
$\dopinion_k=\dopinion_{0,k}\ast\drec_k$ with $\dopinion_k\coloneqq\pushforward{((\eta+\weightopinion^k(1-\eta))x)}\dopinion_0$ and $\drec_k\coloneqq\pushforward{((1-\weightopinion^k)(1-\eta) x)}\drec$.
We now prove that the (type-2) Wasserstein distance between $\dopinion_k$ and $\mu$ decays exponentially to 0. To ease the notation, we write $\dopinion=\bar\dopinion_0\ast\bar\drec$ with 
$\bar\dopinion_0=\pushforward{(\eta x)}\dopinion_0$ and $\bar\drec=\pushforward{((1-\eta) x)}\drec$.
Then,
\begin{align*}
    \wasserstein{2}{\dopinion_k}{\mu}\!
    &=
    \wasserstein{2}{\dopinion_{0,k}\ast \drec_k}{\bar\dopinion_0 \ast \bar\drec}
    \\
    \overset{\heartsuit}&{\leq }
    \wasserstein{2}{\dopinion_{0,k}\ast\drec_k}{\bar\dopinion_0 \ast \drec_k}
    +
    \wasserstein{2}{\bar\dopinion_0 \ast \drec_k}{\bar\dopinion_0\ast\bar\drec}
    \\
    \overset{\spadesuit}&{\leq}
    \wasserstein{2}{\dopinion_{0,k}}{\bar\dopinion_0}
    +
    \wasserstein{2}{\drec_k}{\bar\drec}
    \\
    \overset{\clubsuit}&{\leq}
    \begin{aligned}[t]
    &\sqrt{\int_{\reals}\abs{\left(\eta+\weightopinion^k(1-\eta)\right)\opinion-\eta \opinion}^2\d\dopinion_0(\opinion)}
    \\
    &
    +\!\sqrt{\int_{\reals}\abs{(1-\weightopinion^k)(1-\eta) \rec-(1-\eta) \rec}^2\d\drec(\rec)}
    \end{aligned}
    \\
    &=
    \weightopinion^k(1\!-\!\eta)\!\left(\sqrt{\int_{\reals}\!\abs{\opinion}^2\d\dopinion_0(\opinion)}\!+\!\sqrt{\int_{\reals}\!\abs{\rec}^2\d\drec(\rec)}\right),
\end{align*}
where in $\heartsuit$ we used triangle inequality; in $\spadesuit$ we used \cref{lemma:wasserstein convolution} and $\mu\ast\nu=\nu\ast\mu$; and $\clubsuit$ leverages \cref{lemma:wasserstein pushforward}.
Since $\dopinion_0,\drec\in\Pp{2}{\reals}$ and $\weightopinion\in(0,1)$, we have $\wasserstein{2}{\dopinion_k}{\dopinion}\to 0$ and thus weak convergence in $\Pp{2}{\reals}$~\cite[Theorem 6.9]{villani2009optimal}.
\end{proof}

\begin{proof}[Proof of~\cref{prop:specialcase:continuous exploration}]
At time $k\in\naturals$, the opinion of a user with bias (and thus initial opinion) $\opinion_0\in\reals$ reads  
\begin{equation}\label{eq:always exploration:proof:user}
    \opinion_{k}
    \!=
    \left(\eta+\weightopinion^k(1-\eta)\right)\opinion_0
    +
    (1-\eta)\sum_{l=0}^{k-1}\weightopinion^{k-1-l}(1-\weightopinion)\rec_l, 
\end{equation}
where we proceed as in~\eqref{eq:no exploration:proof:user} and $\eta\coloneqq \weightbias/(1-\weightopinion)\in[0,1)$ (again, the case $\eta=1$ follows separately).
Thus, the opinion distribution at time $k$ is $\dopinion_k=\dopinion_{0,k} \ast \pushforward{((1-\eta)x)}\drec_k$ with 
\begin{equation*}
\begin{aligned}
    \dopinion_{0,k}&\coloneqq \pushforward{\left(\left(\eta+\weightopinion^k(1-\eta)\right)x\right)}\dopinion_0
    \\
    \drec_k&\coloneqq \ast_{l=0}^{k-1} \pushforward{\left(\weightopinion^{k-1-l}(1-\weightopinion)x\right)}\drec,
\end{aligned}
\end{equation*} 
where $\ast_{l=0}^{k-1}\pushforward{(\weightopinion^{k-1-l}(1-\weightopinion)x)}\drec$ is the short-hand notation for $\pushforward{(\weightopinion^{k-1}(1-\weightopinion)x)}\drec\ast\ldots\ast\pushforward{((1-\weightopinion)x)}\drec$.
We now show that $(\dopinion_k)_{k\in\naturals}$, $(\dopinion_{0,k})_{k\in\naturals}$, and $(\drec_k)_{k\in\naturals}$ converge weakly in $\Pp{2}{\reals}$ to some $\bar\dopinion,\bar\dopinion_0,\bar\drec\in\Pp{2}{\reals}$. 
It suffices that all sequences are Cauchy (w.r.t. the Wasserstein distance); convergence then follows from completeness of the Wasserstein space~\cite[Theorem 6.18]{villani2009optimal}.
We start with $(\drec_k)_{k\in\naturals}$. Let $\varepsilon>0$ (w.l.o.g., $\varepsilon<\weightopinion$) and take
    $N_\rho>\log_\weightopinion(\varepsilon(2\frac{1-\weightopinion}{1+\weightopinion}
        \int_{\reals}\abs{\rec}^2\d\drec(\rec))^{-1/2})$,
well-defined since $\drec\in\Pp{2}{\reals}$.
Then, for $m>n>N_\rho$
\begin{align*}
    &\hspace{-0.5cm}\wasserstein{2}{\drec_n}{\drec_m}^2
    \\
    &=
    \wassersteinLetter{2}
    \begin{aligned}[t]
    \big(&\ast_{l=0}^{n-1} \pushforward{\left(\weightopinion^{n-1-l}(1-\weightopinion)x\right)}\drec,
    \\
    &\ast_{l=0}^{m-1} \pushforward{\left(\weightopinion^{m-1-l}(1-\weightopinion)x\right)}\drec\big)^2
    \end{aligned}
    \\
    &=
    \wassersteinLetter{2}
    \begin{aligned}[t]
    \big(&\delta_{0}\ast\big(\ast_{l=0}^{n-1} \pushforward{\left(\weightopinion^{n-1-l}(1-\weightopinion)x\right)}\drec\big),
    \\
    &\big(\ast_{l=0}^{m-n-1} \pushforward{\left(\weightopinion^{m-1-l}(1-\weightopinion)x\right)}\drec\big)
    \\
    &\hspace{1cm}\ast\big(\ast_{l=0}^{n-1} \pushforward{\left(\weightopinion^{n-1-l}(1-\weightopinion)x\right)}\drec\big)
    \big)^2
    \end{aligned}
    \\
    \overset{\heartsuit}&{\leq}
    \wassersteinLarge{2}{\delta_0}{\ast_{l=0}^{m-n-1} \pushforward{\left(\weightopinion^{m-1-l}(1-\weightopinion)x\right)}\drec}^2
    \\
    &=
    \int_{\reals}\abs{x}^2\d\left(\ast_{l=0}^{m-n-1} \pushforward{\left(\weightopinion^{m-1-l}(1-\weightopinion)x\right)}\drec\right)(x)
    \\
    \overset{\spadesuit}&{=}
    \begin{aligned}[t]
    \int_{\reals^{m-n-1}}&\abs{\rec_0+\ldots+\rec_{m-n-1}}^2
    \\
    &\d\left(\pushforward{\left(\weightopinion^{m-1}(1-\weightopinion)x\right)}\drec\right)(\rec_0)\cdots 
    \\
    &\qquad\d\left(\pushforward{\left(\weightopinion^{n}(1-\weightopinion)x\right)}\drec\right)(\rec_{m-n-1})
    \end{aligned}
    \\
    \overset{\clubsuit}&{=}
    \begin{aligned}[t]
    &\int_{\reals^{m-n-1}}|\weightopinion^{m-1}(1-\weightopinion)\rec_0 +\ldots
    \\&+\weightopinion^{n}(1-\weightopinion)\rec_{m-n-1}|^2\d\drec(\rec_0)\cdots \d\drec(\rec_{m-n-1})
    \end{aligned}
    \\
    \overset{\diamondsuit}&{=}
    2(1-\weightopinion)^2
    \begin{aligned}[t]
    &\int_{\reals^{m-n-1}}\weightopinion^{2m-2}\abs{\rec_0}^2+\ldots \\
    &\hspace{-.7cm}+\weightopinion^{2n}\abs{\rec_{m-n-1}}^2\d\drec(\rec_0)\cdots \d\drec(\rec_{m-n-1})
    \end{aligned}
    \\
    &=
    \begin{aligned}[t]
    2(1-\weightopinion)^2\Big(&\weightopinion^{2m-2}\int_{\reals}\abs{\rec_0}^2\d\drec(\rec_0)+\ldots \\
    &+\weightopinion^{2n}\int_{\reals}\abs{\rec_{m-n-1}}^2\d\drec(\rec_{m-n-1})\Big)
    \end{aligned}
    \\
    &=
    2(1-\weightopinion)^2\left(\weightopinion^{2m-2}+\ldots+\weightopinion^{2n}\right)
    \int_{\reals}\abs{\rec}^2\d\drec(\rec)
    \\
    &=
    2(1-\weightopinion)^2\frac{\weightopinion^{2n+1}-\weightopinion^{2m-1}}{1-\weightopinion^2}
    \int_{\reals}\abs{\rec}^2\d\drec(\rec)
    \\
    &\leq 
    2\frac{1-\weightopinion}{1+\weightopinion}\weightopinion^{2N_\rho}
    \int_{\reals}\abs{\rec}^2\d\drec(\rec),
\end{align*}
where in $\heartsuit$ we leveraged~\cref{lemma:wasserstein convolution}, with $\delta_0\ast\mu=\mu$ and $\mu\ast\nu=\nu\ast\mu$; in~$\spadesuit$ we used the definition of convolution;~$\clubsuit$ follows from $\int_{\reals}\phi(x)\d(\pushforward{f}\mu)(x) = \int_{\reals}\phi(f(x))\d\mu(x)$; and in~$\diamondsuit$ we used that $(a+b)^2\leq 2(a^2+b^2)$ for all $a,b\in\reals$.
By definition of $N_\rho$ we therefore conclude
\begin{equation}\label{eq:always exploration:proof:bar cauchy}
    \wasserstein{2}{\drec_n}{\drec_m}
    \leq 
    \weightopinion^{N_\rho}\sqrt{2\frac{1-\weightopinion}{1+\weightopinion}
    \int_{\reals}\abs{\rec}^2\d\drec(\rec)}<\varepsilon.
\end{equation}
Thus, $(\drec_k)_{k\in\naturals}$ is Cauchy. For $(\dopinion_k)_{k\in\naturals}$, let
$
N_{\mu}>\log_\weightopinion(\varepsilon(1-\eta)^{-1}((\int_{\reals}\abs{\opinion}^2\d\dopinion_0(\opinion))^{1/2}+(2\frac{1-\weightopinion}{1+\weightopinion}
        \int_{\reals}\abs{\rec}^2\d\drec(\rec))^{1/2})^{-1})
$
and $m>n>N_\mu$. The distance between $\dopinion_{0,m}$ and $\dopinion_{0,n}$ is
\begin{align}
    &\hspace{-0.3cm}\wasserstein{2}{\dopinion_{0,m}}{\dopinion_{0,n}}^2 \nonumber
    \\
    \overset{\heartsuit}&{\leq}
    \int_{\reals}\abs{(\eta+\weightopinion^m(1-\eta))\opinion\!-\!(\eta+\weightopinion^n(1-\eta))\opinion}^2\d\dopinion_0(\opinion)  \nonumber
    \\
    &=
    (1-\eta)^2(\weightopinion^n-\weightopinion^m)^2\int_{\reals}\abs{\opinion}^2\d\dopinion_0(\opinion) \nonumber
    \\
    &\leq 
    (1-\eta)^2\weightopinion^{2N_\mu}\int_{\reals}\abs{\opinion}^2\d\dopinion_0(\opinion)
    \label{eq:always exploration:proof:barvsnobar}
\end{align}
where in $\heartsuit$ we used~\cref{lemma:wasserstein pushforward}.
Then,
\begin{align*}
    &\hspace{-.1cm}\wasserstein{2}{\dopinion_m}{\dopinion_n}
    \\
    &=
    \wasserstein{2}{\dopinion_{0,m}\ast\pushforward{((1-\eta)x)}\drec_m}{\dopinion_{0,n}\ast\pushforward{((1-\eta)x)}\drec_n}
    \\
    &\leq
    \begin{aligned}[t]
    &\wasserstein{2}{\dopinion_{0,m}\ast\pushforward{((1-\eta)x)}\drec_m}{\dopinion_{0,n}\ast\pushforward{((1-\eta)x)}\drec_m} 
    \\
    &+ \wasserstein{2}{\dopinion_{0,n}\ast\pushforward{((1-\eta)x)}\drec_m}{\dopinion_{0,n}\ast\pushforward{((1-\eta)x)}\drec_n}
    \end{aligned}
    \\
    \overset{\heartsuit}&{\leq}
    \wasserstein{2}{\dopinion_{0,m}}{\dopinion_{0,n}} + \wasserstein{2}{\pushforward{((1-\eta)x)}\drec_m}{\pushforward{((1-\eta)x)}\drec_n}
    \\
    \overset{\spadesuit}&{\leq}
    \wasserstein{2}{\dopinion_{0,m}}{\dopinion_{0,n}} + (1-\eta)\wasserstein{2}{\drec_m}{\drec_n}
    \\
    \overset{\eqref{eq:always exploration:proof:bar cauchy},\eqref{eq:always exploration:proof:barvsnobar}}&{\leq}
    \!\!\weightopinion^N\!(1-\eta)\!\left(\!\!\sqrt{\int_{\reals}\!\abs{\opinion}^2\d\dopinion_0(\opinion)}\!+\!\sqrt{2\frac{1-\weightopinion}{1+\weightopinion}
    \int_{\reals}\!\abs{\rec}^2\d\drec(\rec)}\right)
    \\
    \overset{\clubsuit}&{<}
    \varepsilon,
\end{align*}
where in $\heartsuit$ we used~\cref{lemma:wasserstein convolution} with $\mu\ast\nu=\nu\ast\mu$, in $\spadesuit$ we used \cref{lemma:wasserstein pushforward}, and in $\clubsuit$ follows from the definition of $N_\mu$.
Thus, $(\dopinion_k)_{k\in\naturals}$ is Cauchy too. By completeness of $\Pp{2}{\reals}$ endowed with the Wasserstein distance~\cite[Theorem 6.18]{villani2009optimal}, $(\dopinion_k)_{k\in\naturals}, (\drec_k)_{k\in\naturals}$ converge to some $\bar\dopinion,\bar\drec\in \Pp{2}{\reals}$, respectively. Finally, by~\eqref{eq:always exploration:proof:barvsnobar}, $(\dopinion_{0,k})_{k\in\naturals}$ is Cauchy too, and it converges to $\bar\dopinion_0=\pushforward{(\eta x)}\dopinion_0$: By~\cref{lemma:wasserstein pushforward}
\begin{align*}
    \wasserstein{2}{\dopinion_{0,k}}{\bar\dopinion_0}
    &\leq 
    \sqrt{\int_{\reals}\abs{(\eta+\weightopinion^k(1-\eta))\opinion-\eta \opinion}^2\d\dopinion_0(\opinion)}
    \\
    &=
    \weightopinion^k(1-\eta)\sqrt{\int_{\reals}\abs{\opinion}^2\d\dopinion_0(\opinion})\to 0
\end{align*}
since $\dopinion_0$ has finite second moment.
Then,
\begin{align*}
    &\hspace{-0.3cm}\wasserstein{2}{\dopinion}{\bar\dopinion}
    \\
    &=
    \wasserstein{2}{\bar\dopinion_0\ast\pushforward{((1-\eta)x)}\bar\drec}{\bar\dopinion} 
    \\
    \overset{\heartsuit}&{\leq}
    \begin{aligned}[t]
    &\wasserstein{2}{\bar\dopinion_0\ast\pushforward{((1-\eta)x)}\bar\drec}{\dopinion_{0,k}\ast\pushforward{((1-\eta)x)}\bar\drec} \\
    &+ \wasserstein{2}{\dopinion_{0,k}\ast\pushforward{((1-\eta)x)}\bar\drec}{\dopinion_{0,k}\ast\pushforward{((1-\eta)x)}\drec_k} \\
    &+ \wasserstein{2}{\dopinion_{0,k}\ast\pushforward{((1-\eta)x)}\drec_k}{\bar\dopinion}
    \end{aligned}
    \\
    \overset{\spadesuit}&{\leq}
    \begin{aligned}[t]
    &\wasserstein{2}{\bar\dopinion_0}{\dopinion_{0,k}} + \wasserstein{2}{\pushforward{((1-\eta)x)}\bar\drec}{\pushforward{((1-\eta)x)}\drec_k} \\
    &+ \wasserstein{2}{\dopinion_{k}}{\dopinion}
    \end{aligned}
    \\
    &=
    \wasserstein{2}{\bar\dopinion_0}{\dopinion_{0,k}} + (1-\eta)\wasserstein{2}{\bar\drec}{\drec_k}
    + \wasserstein{2}{\dopinion_{k}}{\dopinion}\to 0.
\end{align*}
where in $\heartsuit$ follows from triangle inequality and in $\heartsuit$ we used \cref{lemma:wasserstein convolution} with $\mu\ast\nu=\nu\ast\mu$ and the definition of $\mu_k$.
Thus, $\wasserstein{2}{\dopinion}{\bar\dopinion}=0$ and so $\bar\dopinion=\dopinion$. The definition of weak convergence in $\Pp{2}{\reals}$ directly yields convergence of the first two moments and, thus, of the variance, which establishes~\eqref{eq:always exploration:mean and variance} (together with the expressions for the pushforward/convolution of the expected value/variance). 
To prove~\eqref{eq:always exploration:almost gaussian}, it now suffices to deploy~\cref{lemma:discounted clt}. Finally, if $\dopinion_0$ and $\drec$ are Gaussian, then $\dopinion_k$ is Gaussian for all $k$. Let $\dopinion$ be Gaussian with the mean and variance in~\eqref{eq:always exploration:mean and variance}. By Gelbrich~\cite{gelbrich1990formula}, 
\begin{equation*}
    \wasserstein{2}{\dopinion_k}{\dopinion}^2\!=\!
    \abs{\expectedValue{\dopinion_k}{x}-\expectedValue{\dopinion}{x}}^2+\abs{\sqrt{\variance(\dopinion_k)}-\sqrt{\variance(\dopinion)}}^2
\end{equation*}
which converges to $0$ as $k\to\infty$.
So, $(\mu_k)_{k\in\naturals}$ converges weakly in $\Pp{2}{\reals}$ to $\dopinion$. This concludes the proof. 
\end{proof} }

\end{document}